\documentclass{Interspeech2024}

\usepackage[usenames,dvipsnames]{color}
\usepackage{cite}

\interspeechcameraready

\title{SynesLM: A Unified Approach for Audio-visual Speech Recognition and Translation via Language Model and Synthetic Data\thanks{* Equal contributions}}

\name[affiliation={1*}]{Yichen}{Lu}
\name[affiliation={1*}]{Jiaqi}{Song}
\name[affiliation={1}]{Xuankai}{Chang}
\name[affiliation={1}]{Hengwei}{Bian}
\name[affiliation={1}]{Soumi}{Maiti}
\name[affiliation={1}]{Shinji}{Watanabe}

\address{
  $^1$Carnegie Mellon University, Pittsburgh, PA, USA}
\email{\{yichenl5, jiaqison, xuankaic\}@andrew.cmu.edu}

\keywords{Audio-Visual Automatic Speech Recognition, Speech Translation, Multimodal Language Model, Multitask}

\DeclareMathOperator{\SynesLM}{\mathrm{SynesLM}}
\DeclareMathOperator{\CrossEntropy}{\mathrm{CE}}

\begin{document}

\maketitle

\begin{abstract}
In this work, we present \textit{SynesLM}, an unified model which can perform three multimodal language understanding tasks: audio-visual automatic speech recognition(AV-ASR) and visual-aided speech/machine translation(VST/VMT). Unlike previous research that focused on lip motion as visual cues for speech signals, our work explores more general visual information within entire frames, such as objects and actions. Additionally, we use synthetic image data to enhance the correlation between image and speech data. We benchmark SynesLM against the How2 dataset, demonstrating performance on par with state-of-the-art (SOTA) models dedicated to AV-ASR while maintaining our multitasking framework. Remarkably, for zero-shot AV-ASR, SynesLM achieved SOTA performance by lowering the Word Error Rate (WER) from 43.4\% to 39.4\% on the VisSpeech Dataset. Furthermore, our results in VST and VMT outperform the previous results, improving the BLEU score to 43.5 from 37.2 for VST, and to 54.8 from 54.4 for VMT.
\end{abstract}

\section{Introduction}
\textit{Synesthesia} is a neurological condition where stimulation of one sensory pathway involuntarily triggers experiences in another, such as perceiving colors when hearing sounds. This phenomenon highlights the complex integration of multisensory inputs in human cognition, essential for comprehending the world through combined audio and visual stimuli~\cite{newell2016multisensory, CHIOU20131750}. Visual cues, for instance, enhance speech recognition and language understanding, aiding in translation tasks. 

Motivated by these insights, we aim to design a unified model for a range of audio-visual tasks that concurrently use audio and visual inputs. Similar to human learning processes, our model benefits from multitask training. Additionally, we found that incorporating pretrained language model weights further improves our model's performance.
\begin{figure}[ht]
\centering
\includegraphics[width=0.45\textwidth]{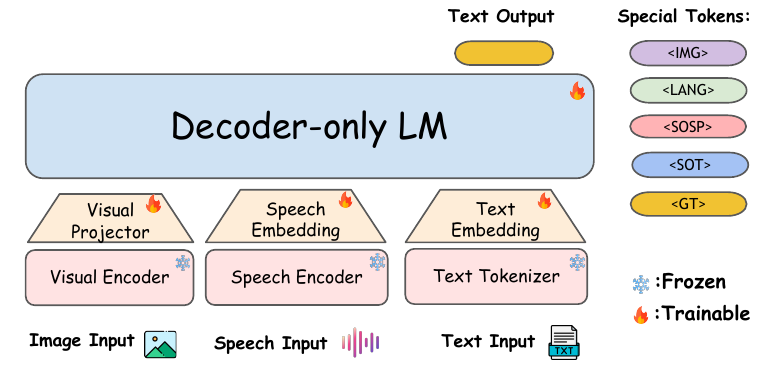} \\
\caption{\textbf{An overview of SynesLM architecture.} The definition of the special tokens will be discussed at the end of the Section~\ref{sec:method}.}
\label{fig:arch}
\end{figure}
Recently, there are several works trying to incorporate visual information with speech to perform automatic speech recognition (ASR). For instance, AV-HuBERT~\cite{av-hubert} focuses on audio-visual speech representation, utilizing video lip recordings to learn powerful speech representations. AVATAR~\cite{AVATAR} extends this concept by incorporating full visual frames for unconstrained audio-visual automatic speech recognition (AV-ASR). The AVFormer~\cite{AVFormer} explores injecting vision into frozen speech models for zero-shot AV-ASR, highlighting the potential of lightweight domain adaptation. However, existing research primarily focuses on AV-ASR, with a noticeable shortage of studies investigating a unified approach for all audio-visual language understanding tasks, such as visual-aided speech translation.

On the other hand, large language models(LLMs) have surged in popularity due to their advanced capabilities in natural language understanding and generation. Several multimodal language models have been developed for more complex tasks necessitating either visual or audio modalities. LLaVa~\cite{llava} and BLIP2~\cite{li2023blip2} corporate visual modality for visual question answering and image captioning. VoxtLM~\cite{voxtlm} and VioLA~\cite{wang2023viola} utilize speech modality to perform multiple speech tasks, such as speech recognition and text-to-speech. Additionally, works like OneLLM~\cite{han2023onellm} and X-LLM~\cite{chen2023xllm} extend the input set to various modalities, including audio and visual. However, none of these models can concurrently process audio and visual inputs, meaning they cannot perform tasks like AV-ASR. 

Since current multimodal language understanding approaches~\cite{AVATAR, AVFormer} are limited to handling AV-ASR, and existing MLLMs can only process tasks involving text plus one additional modality, this paper introduces the \textbf{Synes}thesia \textbf{L}anguage \textbf{M}odel (SynesLM).
SynesLM is a novel, unified approach capable of performing a variety of audio-visual input tasks within a single model. Drawing inspiration from the phenomenon of synesthesia and leveraging advancements in recent language models, SynesLM is meticulously designed to proficiently handle complex language tasks such as audio-visual automatic speech recognition (AV-ASR), visual speech translation (VST), and visual machine translation (VMT). Additionaly, we proposed a data recovery pipeline using LLM and image generative model for multimodal language understanding dataset like How2~\cite{sanabria2018how2} to enhanced its multimodal interaction between speech and visual modality.

We make the following contributions in this work:
\begin{itemize}
    \item \textbf{A unified model for multiple audio-visual tasks:} We explore a novel framework capable of comprehending and processing visual, speech, and textual data to perform various of multimodal language understanding tasks. 
    \item \textbf{Synthetic visual data recovery pipeline:} To address the poor quality of visual data in multimodal speech datasets like How2~\cite{sanabria2018how2}, we developed a novel data recovery pipeline. This pipeline significantly improved the integration between visual and speech data.
    \item \textbf{Performance:} We achieved improved performance across all tasks, recording a 4.0\% WER absolute improvement in zero-shot AV-ASR and BLEU scores of 43.5 for VST and 54.8 for VMT, showcasing our model's strong audio processing and visual comprehension capabilities (Table \ref{table:sota}).
    
    \item \textbf{Reproducibility:} For reproducibility, we open-source our code and model checkpoints in the form of ESPNet recipe in \url{https://github.com/espnet/espnet}.
\end{itemize}

\section{Related Work}

\textbf{Multimodal Language Models (MLMs).} Recently, numerous MLMs have been proposed for various modalities. Advances in Visual Language Models (VLMs), such as~\cite{llava, dai2023instructblip, alayrac2022flamingo, bai2023qwenvl, chen2023shikra, chen2023minigptv2, driess2023palme}, have significantly improved the integration of visual information into pre-trained language models. These models use a pre-trained vision encoder for visual feature extraction, excelling in tasks like image captioning and visual question answering. Similarly, recent studies~\cite{voxtlm, audiolm, audiopalm, dong2023polyvoice, lakhotia2021generative, tang2023salmonn, wu2023speechllama, chen2023salm} have begun exploring unified models for various speech and text tasks, employing self-supervised learning (SSL) feature extractors as audio encoders alongside pre-trained language models to enhance language comprehension. Some recent research~\cite{lyu2023macaw, zhang2023videollama, han2023onellm, chen2023xllm, kim2024tmt} has attempted to incorporate both visual and audio modalities into language models. However, these models can only process one specific modality with text and do not explore the interaction between audio and visual modalities. The most related work to ours is Video-SALMONN~\cite{sun2024videosalmonn}, but it lacks multilingual capabilities, limiting its ability to perform translation tasks.

\noindent\textbf{AV-ASR Methods.} Several methods have been designed for AV-ASR tasks. AV-HuBERT~\cite{av-hubert} uses video recordings to learn robust speech representations through masked multimodal cluster prediction, focusing on lip motion. AVATAR~\cite{AVATAR} employs a multimodal encoder with a transformer decoder for natural language speech recognition output. AVFormer~\cite{AVFormer} explores injecting vision into frozen speech models for zero-shot AVSR, showing the potential for lightweight domain adaptation. Additionally, prompting-whisper~\cite{peng2023whisper} designs a cascade model that injects visual prompts for AV-ASR. However, these methods are limited to AV-ASR tasks or utilize cascade structures, while we design an end-to-end approach for multiple speech-visual tasks.
\section{Method}
\label{sec:method}

\begin{figure}[ht]
\centering
\includegraphics[width=0.45\textwidth]{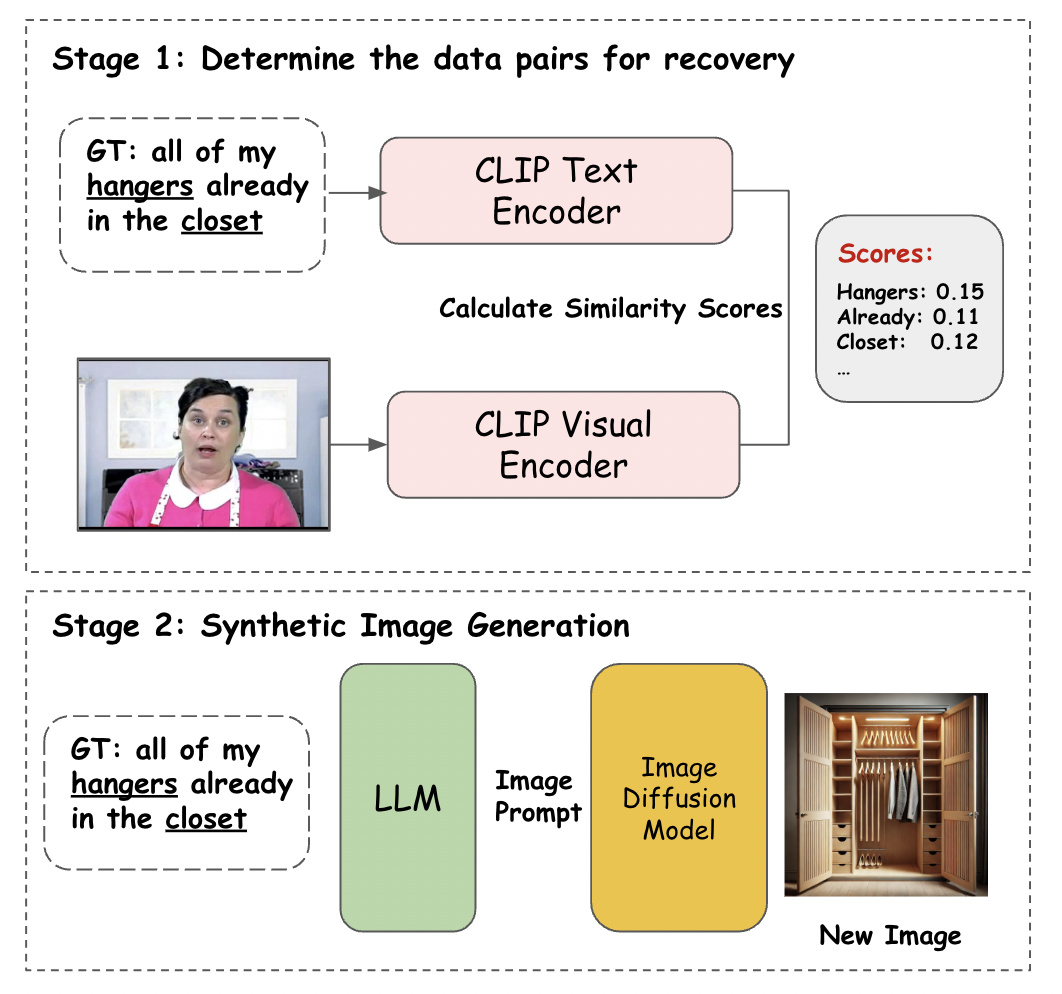} \\
\caption{\textbf{Synthetic Data Recovery Pipeline.} }
\label{fig: pipe}
\end{figure}

\begin{figure*}[ht]
\centering
\includegraphics[width=0.93\textwidth]{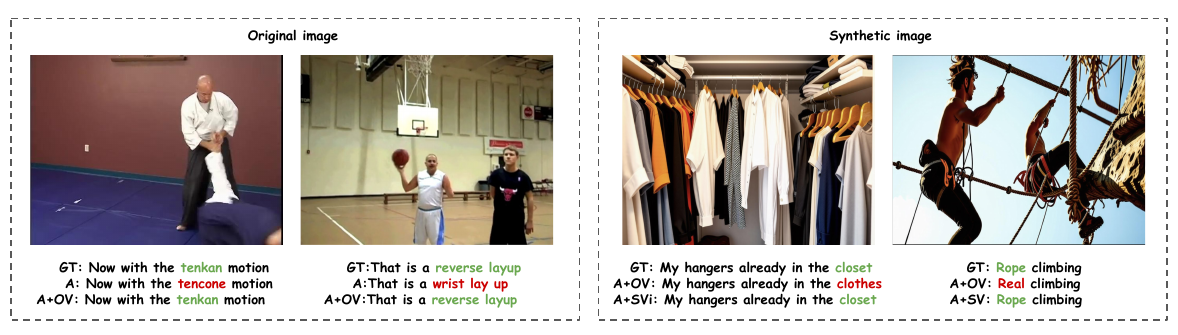} \\
\caption{\textbf{Qualitative examples on How2 ASR.} We show that our audio with original visual (A+OV) and audio with synthetic visual (A+SV) method successfully extract and understand the information from the image and corporate the information with speech representation to perform ASR task.}
\label{fig:example}
\end{figure*}

This section initially outlines the architecture of our model, followed by a description of how we recover the multimodal data and tokenize it.

\subsection{Data Representation}

\textbf{Discrete Speech Representation.} Recent advances in discrete speech representations for ASR and ST have improved training speed, inference speed, and storage efficiency~\cite{dst}. These methods utilize SSL's ability to capture linguistic and acoustic information, surpassing previous techniques like log-Mel filterbanks. By converting continuous audio features into discrete speech tokens, speech input can be handled similarly to text tokens. This unification allows speech and text tokens, sharing similar semantics, to be processed together in a single LM using a unified character set called discrete speech-text tokens.

\noindent\textbf{Visual Encoders and Features.} For the visual modality, we randomly select a single frame from each video clip as the visual input, keeping the image whole rather than dividing it into patches. This simplifies the extraction of object and action information, ensuring one frame provides sufficient details. Using a pre-trained visual encoder from CLIP~\cite{radford2021learning}, we extract features from the entire image. To bridge the visual and language modalities, a Vision-Language connector layer (a Multi-Layer Perceptron) maps these visual features into the same embedding space as the discrete audio-text tokens, aligning them with the linguistic components. We also experiment with other CLIP-like pretrained visual encoders~\cite{zhai2023sigmoid, sun2023evaclip} to explore their performance in fusing discrete text-speech representations.

\noindent\textbf{Data Format.} We use several special tokens to indicate different tasks and different modality. The entire sequence start with a \texttt{<SOS>} token, end with a \texttt{<EOS>} token. We use a \texttt{<IMG>} token to indicate the position of visual information in input sequence. \texttt{<start-of-text>} and \texttt{<start-of-speech>} (\texttt{<SOT>}, \texttt{<SOSP>} in Figure \ref{fig:arch}) represent the different input modalities. \texttt{<generate-text>} (\texttt{<GT>} in Figure \ref{fig:arch}) indicate the output modality of SynesLM. Moreover, the language token \texttt{<LANG>} (e.g. \texttt{<EN>} and \texttt{<PT>}) in the input and output sequence can indicate the source and target language, which enable the translation capacity of our framework. The arrangement of modality and language tokens in SynesLM facilitates its multitasking capability, where each specific combination signifies a distinct task. For example, the token combination \{\texttt{<IMG>}, \texttt{<SOSP>}, \texttt{<EN>}, \texttt{<GT>}, \texttt{<PT>}\} represents the task of VST from English to Portuguese.

\subsection{SynesLM}

Figure \ref{fig:arch} illustrates the overall architecture of SynesLM. We utilize a transformer-based decoder-only LM as our backbone. We use pre-trained OPT model \cite{zhang2022opt} to initialize our weights to achieve better performance and training efficiency \cite{hassid2024textually}. 

To process the speech inputs, SynesLM employs 
a SSL feature extractor like HuBERT \cite{hsu2021hubert} and k-means to generate speech discrete tokens. The integration of visual information is accomplished through a visual encoder and a Vision-Language MLP projector. Details will be explained later in this section.


We concatenate the input text sequence $Y = (y_{i}\in\mathcal{V}_{\text{txt}}|i = 1, ..., t_{txt})$ with the input discrete speech token(dst) sequence $D = (d_{i}\in\mathcal{V}_{\text{dst}}|i = 1, ..., t_{dst})$ as the $T$-length discrete speech-text token sequence $Z = (z_{i}\in\mathcal{V}=\mathcal{V}_{\text{txt}}\cup\mathcal{V}_{\text{dst}}\cup\mathcal{V}_{\text{special}}|i = 1, ..., T)$, where $T=t_{txt}+t_{dst}$ and $\mathcal{V}_{\text{special}}$ is the special token set e.g. \texttt{<LANG>} etc. We feed the input discrete speech-text sequence $Z$ into the embedding layer to get the $D$-dimensional embedding space expressed by $\mathbf{E} = (\mathbf{e}_{i}\in\mathbb{R}^D|i = 1, ..., T)$.
 Additionally, we add a single visual token in the $D$-dimensional embedding space as $\mathbf{V}$.
We feed the visual embedding $\mathbf{V}$ and discrete speech-text embedding $\mathbf{E}$ in to our  auto-regressive LM. The joint probability of the generated text sequence $Y$ can be expressed as:
\begin{equation}
p(Y) = \prod_{i=T+1}^{T+1+L}p \left(Y_{i}|\mathbf{V}, \mathbf{E}_{1:i-1}\right),
\end{equation}
where $L$ is the generated text length.

During training and inference process, given the previous information $\mathbf{V}$, $\mathbf{E}_{1:i-1}$, we can use this formula to predict the probability distribution in the next time step $\hat{p_i}$ by $\hat{p_i} = \SynesLM\left(Y_{i}|\mathbf{V}, \mathbf{E}_{1:i-1}\right)$. Then the cross entropy (CE) loss of the model can be expressed as:
\begin{equation}
 \mathcal{L}_{\CrossEntropy}\left(p_i, \hat{p_i}\right) = - \sum_{c=1}^{|\mathcal{V}|}p_i(c)\log\hat{p_i}(c),
 \end{equation} 
 where $p_i$ is the reference probability distribution.

\subsection{Data Recovery via Synthetic Image Data}

We mainly utilize two different multimodal datasets in our study: 
\begin{itemize}
    \item \textbf{How2}  \cite{sanabria2018how2} is a rich multimodal collection of instructional videos with English subtitles and Portuguese translations. We use the 300 hours subset of video content in this project. We utilize the training set contains 182,167 clips, the validation set includes 1,939 clips, and the test set consists of 2,298 clips. 
    \item \textbf{VisSpeech}  \cite{AVATAR} is constructed by selecting 482 video clips from the extensive HowTo100M dataset  \cite{miech19howto100m}, aimed at creating a robust test set. The selected clips are characterized by their strong visual-audio correlation, providing an ideal platform to evaluate the effectiveness of our visual modality.
\end{itemize}

We use the How2 dataset~\cite{sanabria2018how2} for training and testing. To ensure the correlation between visual and textual data, we initially calculate the similarity score between the ground truth transcription and the selected frame. As illustrated in Figure \ref{fig: pipe}, for a dataset containing image-text pairs, we input the images and text into CLIP~\cite{radford2021learning} to calculate the cosine similarity score for each word in the ground truth. The distribution of similarity scores indicates that over 62\% of visual data have a similarity score below 20\%, reflecting the poor quality of the How2 visual data. To address this issue, we designed a data recovery pipeline to enhance modality interaction within the How2 dataset. If the maximum score for each words in the sentence is below a given threshold ($\tau = 0.2$ in our experiment), we generate new visual data based on the ground truth. For image data generation, we first input the ground truth data into a Large Language Model (LLM) to extract object or action information and generate an image generation prompt based on that information. Finally, we feed the image generation prompt into our image diffusion model~\cite{rombach2022highresolution}.

\section{Experiments}
\label{sec:exp}

\begin{table*}[htbp]
\centering
\caption{Experimental results comparing on single-task and multi-task on different visual encoders. The relative improvement rate represents the performance influence of incorporating visual modality to the original model under the certain tasks, the visual influence in single-task or multi-task scenarios are calculated independently.}
\begin{tabular}{l|llll}
\toprule
                           & \multicolumn{1}{c}{\textbf{ASR (How2)}}        & \multicolumn{1}{c}{\textbf{ASR (VisSpeech)}}   & \multicolumn{1}{c}{\textbf{ST}}                & \multicolumn{1}{c}{\textbf{MT}}                \\
                           & \multicolumn{1}{c}{WER($\downarrow$)} & \multicolumn{1}{c}{WER($\downarrow$)} & \multicolumn{1}{c}{BLEU($\uparrow$)}  & \multicolumn{1}{c}{BLEU($\uparrow$)}  \\ \midrule
                           How2 Baseline w/  visual ~\cite{sanabria2018how2} & 18.0 &    { -} & 37.2 & 54.4 \\ \midrule
Single w/o visual       & 17.6                       & 41.6                       & 40.5                       & 55.2                       \\
Single w/  visual (CLIP \cite{radford2021learning})            & \textbf{17.0  \textcolor{Green}{(+3.41\%)}} & 41.7    (-0.24\%)          & 40.7    \textcolor{Green}{(+0.49\%)}          & \textbf{55.6    \textcolor{Green}{(+0.72\%)}} \\
Single w/  visual (EVA-CLIP \cite{sun2023evaclip})          & 17.6    (+0.00\%)          & \textbf{40.4    \textcolor{Green}{(+2.88\%)}} & 41.3    \textcolor{Green}{(+1.98\%)}          & 54.7    (-0.91\%)          \\
Single w/  visual (SigLIP \cite{zhai2023sigmoid})           & 17.3    \textcolor{Green}{(+1.70\%)}          & 42.3    (-1.68\%)          & \textbf{41.4    \textcolor{Green}{(+2.22\%)}} & 54.7    (-0.91\%)          \\ \midrule
Multi w/o visual  & 16.4                       & 40.8                       & 42.9                       & 54.7                       \\
Multi w/  visual (CLIP \cite{radford2021learning})       & 16.1    \textcolor{Green}{(+1.83\%)}          & 40.1    \textcolor{Green}{(+1.72\%)}          & 43.0    \textcolor{Green}{(+0.23\%)}          & 54.0   (-1.30\%)          \\
Multi w/  visual (EVA-CLIP \cite{sun2023evaclip})  & 15.9    \textcolor{Green}{(+3.05\%)} & 40.2    \textcolor{Green}{(+1.47\%)}          & 43.4    \textcolor{Green}{(+1.17\%)}          & 53.9    (-1.46\%)          \\
Multi w/  visual (SigLIP \cite{zhai2023sigmoid})    & \textbf{15.7    \textcolor{Green}{(+4.27\%)}}          & \textbf{39.4    \textcolor{Green}{(+3.43\%)}} & \textbf{43.5    \textcolor{Green}{(+1.40\%)}} &  \textbf{54.8    \textcolor{Green}{(+0.18\%)}} \\  \bottomrule
\end{tabular}
\label{tabel:visual}
\end{table*}

\begin{table}[h]
\centering
\caption{\textbf{Ablation Study.} Experimental AV-ASR task results comparing the visual influence on multi-task SigLIP\cite{zhai2023sigmoid} encoder. Random visual means randomly select the visual features from other video clips.}
\begin{tabular}{@{}l|cc@{}}
\toprule
        & \begin{tabular}[c]{@{}c@{}}\textbf{ASR (How2)}\\ WER($\downarrow$)\end{tabular} & \begin{tabular}[c]{@{}c@{}}\textbf{ASR (Visspeech)}\\ WER($\downarrow$)\end{tabular} \\ \midrule
Multi w/o visual   & 16.4                                                            & 40.8                                 \\ Multi w random visual & 16.4 & 41.0 
\\ Multi w visual & 15.9 & 39.4 
\\ 
Multi w visual + synthetic & \textbf{15.7}                                                            & \textbf{39.4}                                                       \\ \bottomrule
\end{tabular}
\label{table:ssl}
\end{table}

\subsection{Experiment Setup}
According to the survey of dicrete representation~\cite{dst}, we fixed our BPE size to 3K and k-mean cluster number to 2K since How2 only contains about 300 hours of speech data. We did all of our experiment on the open-source E2E speech processing toolkit ESPNet~\cite{espnet}. For the speech encoder, we select XLS-R~\cite{babu2021xlsr} as our SSL feature extraction method. In all experiments, we use 2$\times$ V100 GPUs for training our models. Our core model is based on a Decoder-only Transformer architecture, featuring 12 layers with a 768-dimensional feature space and 12 attention heads, culminating in a total of 125 million trainable parameters. For multitask training, we ensured an equal distribution of data across the different tasks. We froze the vision encoder to concentrate training on the vision-language projector and the decoder-only language model, achieving an end-to-end training process.

\subsection{Results}
Table \ref{tabel:visual} shows the experimental results evaluating the impact of visual features in both single-task and multi-task settings. Our methodology is evaluated across three distinct tasks: ASR, ST, and MT. Furthermore, we explore the effectiveness of three different visual encoders~\cite{radford2021learning, sun2023evaclip, zhai2023sigmoid} to determine which encoder best aligns with speech-text discrete representations.

There are some notable instances where the visual modality enhances ASR task performance. As demonstrated in Figure \ref{fig:example}, visual cues are instrumental in recognizing seldom-used vocabulary, especially when there is a strong correlation between the visual content and these specific words. This highlights the potential of language models to comprehend visual information and merge it with speech data for a multimodal understanding. The findings indicate that incorporating SynesLM's visual modality consistently enhances performance across all the tasks when compared to an audio-only baseline (e.g., improving the WER from 16.4\% to 15.7\% in a multitask setting using SigLIP \cite{zhai2023sigmoid}). Among the visual encoders tested, CLIP \cite{radford2021learning} achieves the highest performance in single-task experiments. On the other hand, in the multitask framework, SigLIP \cite{zhai2023sigmoid} demonstrates superior efficacy, notably achieving a 3.43\% relative performance increase in the zero-shot AV-ASR on the VisSpeech \cite{AVATAR} dataset. Furthermore, to delve deeper into the impact of visual features, we conduct an ablation study where the visual input is replaced with a random image. As shown in Table \ref{table:ssl}, the WER for the random visual input scenario increases from 15.7\% to 16.4\%, which is the same as the performance without visual input. In addition, our synthetic data recovery technique further improves the performance from 15.9\% to 15.7\%, indicating that better audio-visual correlation could further benefit model performance. This outcome underscores the robustness of our model in audio-visual tasks.

\begin{table}[h]
\centering

\caption{Comparison with the state-of-the-art on AV-ASR task. The \texttt{train set} section for last two rows indicates that those methods use additional dataset other then How2 \cite{sanabria2018how2} for pre-training. \textsuperscript{\textdagger} denotes initialization with OPT. Results are reported as WER (\%, lower is better).}
\begin{tabular}{@{}cc|cc@{}}
\toprule
\textbf{Method}    & \textbf{Train Set}      & \textbf{How2} & \textbf{VisSpeech} \\ \midrule
How2 Base \cite{sanabria2018how2} & 300hrs      & 18.0 & -         \\
LLD \cite{ghorbani2020listen}       & 300hrs      & 16.7 & -         \\
VAT \cite{grounding}       & 300hrs      & 18.0 & -         \\
MultiRes \cite{paraskevopoulos2020multiresolution}  & 300hrs      & 20.5 & -         \\
AVATAR \cite{AVATAR}    & 300hrs      & \textbf{15.6} & 43.4      \\
\textbf{Ours\textsuperscript{\textdagger}}      & 300hrs      & 15.7 & \textbf{39.4}      \\ \midrule
AVFormer \cite{AVFormer}  & 960hrs + 6500hrs     & 13.6 & 16.6      \\
AVATAR \cite{AVATAR}    & 300hrs + 131k hrs & 9.1  & 11.3     \\
Prompt-whisper \cite{peng2023whisper}    & 680k hrs & -  & 7.16     \\
\bottomrule
\end{tabular}
\label{table:sota}
\end{table}

\noindent\textbf{Compare with the SOTA methods.} 
In Table \ref{table:sota}, we compare our model with state-of-the-art methods on the AV-ASR task. The results demonstrate that our approach surpasses most of the methods when utilizing only the How2 dataset. Note that, both AVATAR \cite{AVATAR} and AVFormer \cite{AVFormer} pretrain on the vast HowTo100M \cite{miech19howto100m} dataset, which makes them perform well explicitly on single AV-ASR task. Contrary to the design purposes of these models, we aim to explore an unified model architecture for different audio-visual related tasks. Under the multitasking scenario, our model not only retains high performance in the AV-ASR task but also surpasses the How2 baseline in ST and MT tasks. Specifically, as shown in Table \ref{tabel:visual}, we observed a significant improvement in the BLEU score for ST, increasing from 37.2 to 43.5. Similarly, for MT, there was a modest enhancement from 54.4 to 54.8 in the BLEU score.



\section{Conclusion}
We introduce SynesLM, a novel multimodal language model designed for multiple audio-visual tasks. SynesLM outperforms existing single-task methods, showcasing the effective integration of audio and visual data. The experiments also reveal SynesLM's ability to synergize auditory, textual and visual information effectively. The results underscore SynesLM's proficiency across all evaluated tasks and highlight its potential for broader applications in audio-visual processing. 

\section{Acknowledgements}
\ifinterspeechfinal
Experiments of this work used the Bridges2 system at PSC  and Delta system at NCSA through allocations CIS210014 and IRI120008P from the Advanced Cyber infrastructure Coordination Ecosystem: Services \& Support (ACCESS) program, supported by National Science 
Foundation grants \#2138259, \#tel:2138286, \#tel:2138307, \#tel:2137603, and \#tel:2138296.
\else

\fi

\footnotesize
\bibliographystyle{IEEEtran}
\bibliography{mybib}

\end{document}